\documentclass[aps,prb]{revtex4}

\usepackage{amsmath}
\usepackage{amssymb}
\usepackage{epsfig}

\newcommand {\dr}{{\mathrm d}\mathbf{r}}

\newcommand {\dk}{{\mathrm d}\mathbf{k}}
\newcommand {\dpp}{{\mathrm d}\mathbf{p}}
\newcommand {\dd}{{\mathrm d}}
\newcommand {\rr}{\mathbf{r}}
\newcommand {\kk}{\mathbf{k}}

\newcommand {\pp}{\mathbf{p}}
\newcommand {\X}{\mathbf{X}}

\newcommand {\F}{\mathbf{F}}
\newcommand {\jj}{\mathbf{j}}
\newcommand {\1}{\mathbf{1}}

\newcommand {\vv}{{\mathbf v}}
\newcommand {\LL}{{\mathbf \Lambda}}
\newcommand {\AAA}{{\mathbf A}}
\newcommand {\FF}{\mathbf{F}}

\begin{document}

\title{Dynamical density functional theory for molecular and colloidal fluids:
a microscopic approach to fluid mechanics}

\author{A.J. Archer}
\affiliation{Department of Mathematical Sciences,
Loughborough University,
Loughborough,
Leicestershire,
LE11 3TU, UK}

\date{\today}

\begin{abstract}
In recent years, a number of dynamical density functional theories (DDFTs) have been developed for describing the dynamics of the one-body density of both colloidal and atomic fluids. In the colloidal case, the particles are assumed to have stochastic equations of motion and theories exist for both the case when the particle motion is over-damped and also in the regime where inertial effects are relevant. In this paper we extend the theory and explore the connections between the microscopic DDFT and the equations of motion from continuum fluid mechanics. In particular, starting from the Kramers equation which governs the dynamics of the phase space probability distribution function for the system, we show that one may obtain an approximate DDFT that is a generalisation of the Euler equation. This DDFT is capable of describing the dynamics of the fluid density profile down to the scale of the individual particles. As with previous DDFTs, the dynamical equations require as input the Helmholtz free energy functional from equilibrium density functional theory (DFT). For an equilibrium system, the theory predicts the same fluid one-body density profile as one would obtain from DFT. Making further approximations, we show that the theory may be used to obtain the mode coupling theory that is widely used for describing the transition from a liquid to a glassy state.
\end{abstract}


\maketitle

\section{Introduction}
\label{sec:intro}

To study fluid dynamical phenomena it is often sufficient to consider the
fluid as a continuum and ignore the fact that it is in reality made up of
individual particles. For example, such an approach leads to the Navier-Stokes
equation, a corner stone in the field of Fluid Dynamics.\cite{Kreuzer, Acheson, Batchelor}
It is only if one is
interested in the dynamical behaviour on length scales comparable with the size
of the individual particles that one must go beyond a continuum theory.
This is rarely the case for atomic and molecular fluids. However, colloidal
suspensions are a particular class of fluids where,
due to the mesoscopic size of the colloids, this limit is more easily reached.
For example, to describe blood flow in the capillaries, due to the fact that
the diameter of the red blood cells can be similar to that of the
capillaries, one can not treat the fluid (blood) as a continuum. This regime is
also reached in a number of microfluidic devices\cite{microfluidics} and in laser tweezer experiments,\cite{tweezers} where
groups or individual colloids are trapped and moved around by the tweezer.

Over the last few years a dynamical density functional theory (DDFT)
has been developed. This constitutes a microscopic theory for the fluid dynamics
of such colloidal fluids. The starting point was work by
Marconi and Tarazona,\cite{MT,MT2} in which they assumed that the colloids can
be modelled as Brownian particles with stochastic equations of motion, thus
neglecting hydrodynamic interactions between the colloids.
Newton's equations of motion for a system of $N$ identical Brownian particles of
mass $m$, specified by the set
of particle position coordinates $\rr^N=\{\rr_1,\rr_2,\cdots,\rr_N\}$ and
momenta $\pp^N=\{\pp_1,\pp_2,\cdots,\pp_N\}$, are:\cite{Dhont_book}
\begin{eqnarray}
\frac{d \rr_i}{dt}=\frac{\pp_i}{m}, \nonumber \\
\frac{d \pp_i}{dt}=-\gamma \pp_i+\X_i(\rr_i)+ {\bf G}_i(t),
\label{eq:eqs_of_motion}
\end{eqnarray}
where the forces due to the solvent on the colloidal particles are modelled solely by a
viscous drag term, $-\gamma \pp_i$, where $\gamma$ is a friction coefficient,
and by ${\bf G}_i(t)=(\xi_i^x(t),\xi_i^y(t),\xi_i^z(t))$, a stochastic
white noise term
with the property $\left< \xi_i^{\alpha}(t) \right> =0$ and
$\left< \xi_i^{\alpha}(t)\xi_i^{\nu}(t')\right> = 2 \gamma m k_BT \delta_{ij}
\delta^{\alpha \nu} \delta(t-t')$, where $k_B$ is Boltzmann's constant and $T$
is the temperature. $\X_i$ is the sum of the force on particle $i$ due to any external potentials and the forces due to interactions with the other particles in the system, and is given by $\X_i=-\nabla_{\rr_i}V(\rr^N,t)$, where the potential energy $V(\rr^N,t)$ is assumed to be of the form
\begin{eqnarray}
V(\rr^N,t)=&\sum_iV^{ext}(\rr_i,t)
+\frac{1}{2}\sum_{i,j}v_2(\rr_i(t),\rr_j(t)) \nonumber \\
&+\frac{1}{6}\sum_{i,j,k}v_3(\rr_i(t),\rr_j(t),\rr_k(t))+ \cdots,
\label{eq:PE_3bod}
\end{eqnarray}
where $V^{ext}(\rr,t)$ is the external one-body potential, $v_2$ is the pair interaction between the particles, $v_3$ is the triplet interaction (which is often assumed to be zero) and there may also be other higher body terms,\cite{Archer7,Archer11} which are denoted by $\cdots$ in Eq.\ \eqref{eq:PE_3bod}. We should emphasise that the hydrodynamic interactions between the colloidal particles, which can be vital for a quantitative description of the dynamical processes in dense suspensions, have been neglected in Eq.\ \eqref{eq:eqs_of_motion}. To include these effects, the term in Eq.\ \eqref{eq:eqs_of_motion} describing the force exerted by the solvent on colloid $i$, $\FF_i^h=-\gamma\pp_i$, should be replaced by $\FF_i^h=-\sum_{j=1}^N\Upsilon_{ij}(\rr^N)\cdot\pp_j$, where $\Upsilon_{ij}(\rr^N)$ is a microscopic friction matrix that depends on the coordinates of {\em all} the particles.\cite{Dhont_book}

There are two limits one may consider in Eq.\ \eqref{eq:eqs_of_motion}. The first is the limit when $\gamma$ is small. In the limit $\gamma \to 0$, both the viscous drag term and the stochastic noise term in Eqs.\ \eqref{eq:eqs_of_motion} disappear (recall that they are related to one another by a fluctuation-dissipation relation) and Eq.\ \eqref{eq:eqs_of_motion} reduces to a (deterministic) set of Newton's equations for the motion for a system of particles that are {\em not} immersed in a solvent:
\begin{eqnarray}
 m \frac{d^2 \rr_i}{dt^2} =\X_i(\rr_i).
\label{eq:eqs_of_motion_not_damped}
\end{eqnarray}
These are the equations of motion for an atomic or molecular fluid.

The other limit to consider is when the friction coefficient $\gamma$ is large (overdamped dynamics). In this case the term $d \pp_i/dt$ in Eq.\ \eqref{eq:eqs_of_motion} is small and one may neglect it, so that Eq.~(\ref{eq:eqs_of_motion}) reduces to the following set of equations of motion for the colloids:
\begin{eqnarray}
\gamma m \frac{d \rr_i}{dt} =\X_i(\rr_i)+ {\bf G}_i(t).
\label{eq:eqs_of_motion_damped}
\end{eqnarray}
For a system with these underlying equations of motion, Marconi and Tarazona
obtained the following equation of motion for the one-body density
$\rho(\rr,t)$ of the colloidal particles:\cite{MT, MT2, footnote1}
\begin{equation}
\frac{\partial \rho(\rr,t)}{\partial t}=
\frac{1}{\gamma m} \nabla \cdot \left[ \rho(\rr,t) \nabla \frac{\delta
F[\rho(\rr,t)]}{\delta \rho(\rr,t)} \right],
\label{eq:MT_DDFT_a}
\end{equation}
by making the assumption that the two-body correlations in the non-equilibrium fluid can be approximated by those of an equilibrium fluid with the same one-body density profile.\cite{MT} In Eq.\ \eqref{eq:MT_DDFT_a}, $F[\rho]$ is the {\em equilibrium} Helmholtz free energy functional from the well-established classical density functional theory.\cite{Evans92,HM} It is the fact that the theory builds upon the equilibrium free energy functional that makes the theory so appealing. It means that so long as one has a reliable expression for $F[\rho]$ (and many such approximations are known -- see, for example Refs.\ \onlinecite{Evans92}, \onlinecite{HM} and references therein) then Eq.~(\ref{eq:MT_DDFT_a}) at least guarantees the correct equilibrium profile,\cite{MT,MT2,Archer7} and in practise, it often proves to be reliable out of equilibrium as well.\cite{MT, MT2, joe:christos, Archer7, flor2, flor1, Archer11, Archer13, RexetalPRE2005, RexetalMolP2006, Matthiasetal}

The main reason that one can obtain such a relatively simple expression as Eq.~(\ref{eq:MT_DDFT_a}) for the equation governing the fluid dynamics is because the colloids are suspended in a solvent, which acts as a heat bath, so there are no thermal gradients and one does not explicitly have to take into account conservation of energy. Neither does one have to account for momentum currents, due to the overdamped dynamics. Recall that such considerations are required to obtain reliable dynamical  equations for atomic or molecular fluids such as the Navier-Stokes equation.\cite{Kreuzer} However, even for colloidal fluids one should take account of momentum currents, since strictly speaking, Eq.~(\ref{eq:eqs_of_motion}) provides a better account than Eq.~(\ref{eq:eqs_of_motion_damped}) of the particle dynamics. It was to address this issue that Marconi and Tarazona published a further paper,\cite{MT3} in which they derived, using a multiple time scale analysis, a DDFT that incorporates inertial effects. In the limit where $\gamma$ is large, their theory reduces to their original theory,\cite{MT} i.e.\ where the dynamics is governed by  Eq.~(\ref{eq:MT_DDFT_a}). Their analysis generates a hierarchy of coupled equations which in principle must be solved self consistently. They also argue that so long as $\gamma$ is not too small, then one can describe the fluid dynamics using just the truncation of the expansion at second order.\cite{MT3, MarconiMelchionna}

In this paper we take a different route and derive an alternative DDFT, applicable to the same kinds of systems. We do not use the multiple time scale analysis that Marconi and Tarazona employed.\cite{MT3} Instead, starting from the Kramers equation for the $N$-particle phase space probability density distribution function of the system $f^{(N)}$, we derive an equation (Eq.\ \eqref{eq:DDFT_1}), which describes the dynamics of the momentum currents $\jj$ in the fluid of Brownian particles. In this exact equation, we make two approximations: (i) We assume that one can approximate the two body spatial correlations in the non-equilibrium fluid by those of an equilibrium fluid with the same one body density profile -- this is also the approximation used to obtain Eq.~(\ref{eq:MT_DDFT_a}). (ii) We assume a local equilibrium Maxwell-Boltzmann form for the one particle phase space probability density distribution function $f^{(1)}$. From these two approximations we are able to derive a DDFT that takes the form of a generalised Euler equation. The Euler equation may be obtained from the Navier-Stokes equation by setting the shear and the bulk viscosity to zero. The same is true for the present system: assumption (ii) above is equivalent to setting these viscosities to zero, and following Kreuzer\cite{Kreuzer} we see that by going beyond the local equilibrium approximation for $f^{(1)}$, we may obtain a DDFT that takes the form of a generalised Navier-Stokes equation. These two equations from fluid mechanics are very well known, and provide the starting point for the description of many fluid dynamical phenomena. A large body of knowledge has been built up in the literature concerning solutions of the Navier-Stokes and other such equations from fluid dynamics.\cite{Kreuzer, Acheson, Batchelor} The contribution of the present paper is to show how to connect these continuum theories with the fully microscopic DDFT, and to indicate how one may incorporate information about the microscopic structure and correlations of the fluid into the continuum theories.

To solve both the generalised Euler and generalised Navier-Stokes DDFTs mentioned above, requires explicitly keeping track of both the fluid density $\rho(\rr,t)$ and the local fluid velocity $\vv(\rr,t)$. However, by neglecting certain terms in the dynamical equations, we show that one may obtain a DDFT that only explicitly depends of the fluid density $\rho(\rr,t)$. From this DDFT, following the approach given in Ref.\ \onlinecite{Archer17}, we are able to obtain a mode coupling theory (MCT) for the density fluctuation correlation function, that is of the standard MCT form.

This paper is laid out as follows: In Sec.~\ref{sec:theory}, we derive a DDFT for the average one-body density $\rho(\rr,t)$ of a fluid of colloidal particles, whose equations of motion are given by Eq.~(\ref{eq:eqs_of_motion}). In Sec.~\ref{sec:MCT} we use our DDFT to obtain the MCT. Finally, in Sec.~\ref{sec:conc} we summarise and draw some conclusions.

\section{Equations of motion}
\label{sec:theory}

For a system of $N$ identical particles, whose equations of motion are given by Eq.~(\ref{eq:eqs_of_motion}), the time evolution of the phase space probability density function, $f^{(N)}(\rr^N,\pp^N,t)$, which gives the probability of the system being in a particular configuration $(\rr^N, \pp^N)$ at time $t$, is governed by the Kramers (Fokker--Planck) equation:\cite{vanKampen}
\begin{eqnarray}
\frac{\partial f^{(N)}}{\partial t}
+\frac{1}{m} \sum_{i=1}^N \pp_i \cdot \nabla_{\rr_i} f^{(N)}
+\sum_{i=1}^N \X_i \cdot \nabla_{\pp_i} f^{(N)}
= \gamma \sum_{i=1}^N \nabla_{\pp_i} \cdot \pp_i f^{(N)}
+ \gamma m k_BT \sum_{i=1}^N \nabla_{\pp_i}^2 f^{(N)}.
\label{eq:kramers}
\end{eqnarray}
Note that in the limit $\gamma \to 0$ this reduces to the Liouville equation,\cite{HM} which is the equation governing the time evolution of $f^{(N)}$ when the system equations of motion are given by Eq.\ \eqref{eq:eqs_of_motion_not_damped}. We may define a set of reduced phase space distribution functions:\cite{HM}
\begin{widetext}
\begin{eqnarray}
f^{(n)}(\rr^n,\pp^n,t)=\frac{N!}{(N-n)!}\int \dr^{(N-n)}
\int \dpp^{(N-n)} f^{(N)}(\rr^N,\pp^N,t).
\label{eq:f_reduced}
\end{eqnarray}
Integrating over the Kramers equation \eqref{eq:kramers}, we obtain the following dynamical equation for the one particle reduced phase space distribution function $f^{(1)}$:
\begin{eqnarray}
\left(\frac{\partial}{\partial t} + \frac{\pp_1}{m}\cdot\nabla_{\rr_1}
+ \F^{ext}_1\cdot\nabla_{\pp_1} \right) f^{(1)}(\rr_1,\pp_1,t)
=\gamma \left[\nabla_{\pp_1}\cdot \pp_1+mk_BT\nabla_{\pp_1}^2 \right]
f^{(1)}(\rr_1,\pp_1,t) \nonumber \\
-\int \dr_2\int \dpp_2 \F_{12} \cdot \nabla_{\pp_1}
f^{(2)}(\rr_1,\pp_1,\rr_2,\pp_2,t)+\cdots,
\label{eq:BBGKY_1}
\end{eqnarray}
\end{widetext}
where $f^{(2)}$ is the two-particle distribution function, $\F^{ext}_1=-\nabla_{\rr_1}V^{ext}(\rr_1)$, $\F_{12}=-\nabla_{\rr_1}v_2(\rr_1-\rr_2)$ and $\cdots$ contains contributions from three-body and higher-body interactions. We have also assumed that all boundary terms are zero -- i.e.\ that $f^{(n)}(\rr^n,\pp^n,t)$ is zero when any of the components of $\rr_i$ and $\pp_i \to \pm \infty$. Note that when $\gamma=0$, Eq.\ \eqref{eq:BBGKY_1} is simply the first equation in the Bogolyubov-Born-Green-Kirkwood-Yvon (BBGKY) hierarchy.\cite{HM} The following manipulations are similar to those in Refs.\ \onlinecite{Kreuzer} and \onlinecite{Archer17}, although the starting
point, Eq.~(\ref{eq:BBGKY_1}), is different. In Eq.~(\ref{eq:BBGKY_1}) we may integrate over the momentum $\pp_1$, in order to obtain the continuity equation:
\begin{equation}
\frac{\partial \rho(\rr_1,t)}{\partial t}+\nabla_{\rr_1} \cdot \jj=0,
\label{eq:continuity}
\end{equation}
where the one-body (number) density
\begin{equation}
\rho(\rr_1,t) =\int \dpp_1 f^{(1)}(\rr_1,\pp_1,t),
\end{equation}
and
\begin{equation}
\jj(\rr_1,t) =\int \dpp_1 \frac{\pp_1}{m} f^{(1)}(\rr_1,\pp_1,t)
\label{eq:current}
\end{equation}
is the current.
Taking Eq.~(\ref{eq:BBGKY_1}), multiplying through by $\pp_1/m$ and then integrating over the resultant equation with respect to $\pp_1$, yields
the following:
\begin{widetext}
\begin{eqnarray}
\frac{\partial \jj(\rr_1,t)}{\partial t}+\gamma \jj(\rr_1,t)
+ \nabla_{\rr_1} \cdot \int \dpp_1 \frac{\pp_1 \otimes \pp_1}{m^2}
f^{(1)}(\rr_1,\pp_1,t)
-\frac{1}{m} \rho(\rr_1,t) \X_1 \nonumber \\
-\frac{1}{m}\int \dr_2 \F_{12} \rho^{(2)}(\rr_1,\rr_2,t)
-\frac{1}{m} \int \dr_2 \int \dr_3 \F_{123}
\rho^{(3)}(\rr_1,\rr_2,\rr,_3,t)+\cdots=0,
\label{eq:mom_bal_1}
\end{eqnarray}
where $\F_{123}=-\nabla_{\rr_1}v_3(\rr_1,\rr_2,\rr_3)$ is the three--body force on particle 1 due to particles 2 and 3,
\begin{equation}
\rho^{(2)}(\rr_1,\rr_2,t)
=\int \dpp_1 \int \dpp_2 f^{(2)}(\rr_1,\pp_1,\rr_2,\pp_2,t)
\end{equation}
is the two body density distribution function and
\begin{equation}
\rho^{(3)}(\rr_1,\rr_2,\rr_3,t)
=\int \dpp_1 \int \dpp_2 \int \dpp_3 f^{(3)}(\rr_1,\pp_1,\rr_2,\pp_2,\pp_3,t)
\end{equation}
is the three body density distribution function. Note that $\pp_1 \otimes \pp_1$ in Eq.~(\ref{eq:mom_bal_1}) denotes a tensor product (dyadic). Eq.~(\ref{eq:continuity}), the continuity equation, is simply a statement of conservation of mass and Eq.~(\ref{eq:mom_bal_1}) is a momentum balance equation. These two equations are just Eqs.~(19) and (20) in Ref.\ \onlinecite{MT3}, and it is from here that we take a different approach to the authors of Ref.\ \onlinecite{MT3} in developing the theory.

At equilibrium, the one particle distribution function takes the Maxwell-Boltzmann form:\cite{HM}
\begin{equation}
f^{(1)}(\rr,\pp)=\frac{\rho(\rr)}{(2 \pi m k_BT)^{3/2}}
\exp \left(-\frac{\pp^2}{2 m k_BT}\right).
\label{eq:MB}
\end{equation}
Thus, at equilibrium, we find that the integral $\int \dpp_1 (\pp_1 \otimes \pp_1) f^{(1)}=m k_B T \rho(\rr_1) \1$, where $\1$ denotes the $3\times 3$ unit matrix. Noting this result, we may recast Eq.~(\ref{eq:mom_bal_1}) as follows
\begin{eqnarray}
\frac{\partial \jj(\rr_1,t)}{\partial t}
+\gamma \jj(\rr_1,t)+\AAA(\rr_1,t)+\frac{k_BT}{m}\nabla\rho(\rr_1,t)
+ \frac{1}{m}\rho(\rr_1,t) \nabla V^{ext}(\rr_1,t) \nonumber \\
+\frac{1}{m}\int \dr_2 \rho^{(2)}(\rr_1,\rr_2,t)\nabla v_2(\rr_1-\rr_2) \nonumber \\
+\frac{1}{m} \int \dr_2 \int \dr_3 \rho^{(3)}(\rr_1,\rr_2,\rr_3,t)\nabla v_3(\rr_1,\rr_2,\rr_3) +\cdots=0,
\label{eq:DDFT_1}
\end{eqnarray}
where
\begin{eqnarray}
\AAA(\rr,t)=\nabla \cdot \int \dpp\left(\frac{\pp \otimes \pp}{m^2}-\frac{k_BT}{m} \1 \right)
f^{(1)}(\rr,\pp,t).
\label{eq:B}
\end{eqnarray}
\end{widetext}
From equipartition, it can easily be seen that at equilibrium $\AAA(\rr,t)=0$. Note that in the limit $\gamma \to 0$, i.e.\ when the solvent is absent, Eq.~(\ref{eq:DDFT_1}) is still correct, even though the underlying equations of motion (\ref{eq:eqs_of_motion}) become deterministic.

So far, we have made no approximations and together Eqs.~(\ref{eq:DDFT_1}) and (\ref{eq:B}) are exact. It is at this stage in the derivation that we make two approximations. The first is to assume that we may approximate the two-body spatial correlations in the non-equilibrium fluid by those of an equilibrium fluid with the same one body density profile. This is done by assuming that one can apply the following sum rule (which is exact for the equilibrium fluid) as an approximation for the sixth and seventh terms on the left hand side of Eq.~(\ref{eq:DDFT_1}):\cite{MT2,Archer7}
\begin{eqnarray}
-k_BT \rho(\rr_1) \nabla c^{(1)}(\rr_1)
=\int \dr_2 \rho^{(2)}(\rr_1,\rr_2) \nabla_{\rr_1} v_2(\rr_1,\rr_2) \nonumber \\
+\int \dr_2 \int \dr_3
\rho^{(3)}(\rr_1,\rr_2,\rr_3) \nabla_{\rr_1} v_3(\rr_1,\rr_2,\rr_3)
+ \cdots,
\label{eq:grad_c1_many}
\end{eqnarray}
where $c^{(1)}(\rr)$ is the one body direct correlation function and is equal to the functional derivative of the excess part of the Helmholtz free energy functional:\cite{Evans92,HM}
\begin{equation}
c^{(1)}(\rr) \, = \, -\beta
\frac{\delta F_{ex}[\rho(\rr)]}{\delta \rho(\rr)},
\label{eq:c1}
\end{equation}
where $\beta=1/k_BT$. 

Assuming that we may apply Eq.\ \eqref{eq:grad_c1_many} for the non-equilibrium fluid should be reliable for the case when the particles interact via potentials that are slowly varying.\cite{Archer7, MarconiMelchionna} However, in cases such as a system of hard spheres, where the collision dynamics are somewhat different than in systems of particles interacting via potentials that vary continuously, this approximation is less reliable, and one should approximate this quantity using a term involving a binary collision operator, along the lines described in Refs.\ \onlinecite{MT3} and \onlinecite{MarconiMelchionna}.

Making this approximation in Eq.\ \eqref{eq:DDFT_1}, and also noting that $\nabla \rho=\rho \nabla \ln (\rho)$, we obtain:
\begin{equation}
\frac{\partial \jj(\rr,t)}{\partial t}
+\gamma\jj(\rr,t)+\AAA(\rr,t)+\frac{1}{m}  \rho(\rr,t) \nabla \frac{\delta
F[\rho(\rr,t)]}{\delta \rho(\rr,t)} =0,
\label{eq:mainres}
\end{equation}
where the Helmholtz free energy functional\cite{Evans92,HM}
\begin{eqnarray}
F[\rho]=k_BT \int \dr \rho(\rr)[\ln \Lambda^3 \rho(\rr)-1]
+F_{ex}[\rho] 
+\int \dr V^{ext}(\rr)\rho(\rr).
\label{eq:F}
\end{eqnarray}
The first term on the right hand side is the ideal-gas contribution to the free energy, $\Lambda$ is the thermal de Brogle wavelength and $F_{ex}[\rho]$ is the excess (over ideal-gas) contribution due to interactions between the particles.

We now make a second approximation and we assume that we can make a `local-equilibrium' Maxwell-Boltzmann approximation for the one-particle distribution
function\cite{Kreuzer, HM86} [c.f. Eq.~(\ref{eq:MB})]:
\begin{equation}
f^{(1)}_{l.e.}(\rr,\pp,t)=\frac{\rho(\rr,t)}{(2 \pi m k_BT)^{3/2}}
\exp \left(-\frac{[\pp-\bar{\pp}(\rr,t)]^2}{2 m k_BT}\right),
\label{eq:MB_LE}
\end{equation}
where $\bar{\pp}(\rr,t)=m \vv(\rr,t)$, and $\vv(\rr,t)$ is the average local velocity of the particles. If one substitutes Eq.~(\ref{eq:MB_LE}) into Eq.~(\ref{eq:current}) then one
obtains the following expression for the current:
\begin{eqnarray}
\jj_{l.e.}(\rr,t)=\rho(\rr,t) \bar{\pp}(\rr,t)/m \nonumber \\
=\rho(\rr,t) \vv(\rr,t).
\label{eq:j_le}
\end{eqnarray}
Using this approximation for the current, the continuity equation \eqref{eq:continuity} becomes:
\begin{equation}
\frac{\partial \rho}{\partial t}+\nabla \cdot ( \rho \vv) =0.
\label{eq:xx_2}
\end{equation}
On substituting Eq.~(\ref{eq:MB_LE}) into Eq.~(\ref{eq:B}), we obtain the result
\begin{eqnarray}
\AAA_{l.e.}(\rr,t)=\nabla \cdot \left[\frac{\rho(\rr,t)}{m^2} \bar{\pp}(\rr,t) \otimes \bar{\pp}(\rr,t)\right].
\label{eq:B_LE}
\end{eqnarray}
This, combined with Eq.\ (\ref{eq:mainres}), gives:
\begin{equation}
\frac{\partial (\rho\vv)}{\partial t}
+\gamma\rho\vv+\nabla \cdot (\rho \vv \otimes \vv)=-\frac{1}{m} \rho \nabla \frac{\delta F[\rho]}{\delta \rho}.
\label{eq:xx}
\end{equation}
This expression may be simplified by noting the following results:
\begin{eqnarray}
\nabla \cdot (\rho \vv \otimes \vv)
&=&\rho[ \nabla \cdot (\vv \otimes \vv)]+(\nabla \rho)\cdot(\vv \otimes \vv) \notag \\
&=&\rho[ \vv (\nabla \cdot \vv)+\vv.\nabla\vv]+\vv(\vv \cdot \nabla\rho)
\label{eq:equality_1}
\end{eqnarray}
where $\nabla \vv$ is the tensor derivative, and
\begin{eqnarray}
\frac{\partial (\rho\vv)}{\partial t}
&=&\rho \frac{\partial \vv}{\partial t}+\vv \frac{\partial \rho}{\partial t} \notag \\
&=&\rho \frac{\partial \vv}{\partial t}-\vv [\nabla \cdot (\rho \vv)] \notag \\
&=&\rho \frac{\partial \vv}{\partial t}-\rho \vv (\nabla \cdot \vv)-\vv(\vv \cdot \nabla \rho) 
\label{eq:equality_2}
\end{eqnarray}
where we have used Eq.\ \eqref{eq:xx_2} to obtain the second line in Eq. (\ref{eq:equality_2}). We now use Eqs.\ (\ref{eq:equality_1}) and (\ref{eq:equality_2}) to simplify Eq.\ (\ref{eq:xx}), giving:
\begin{equation}
\rho\left(\frac{\partial \vv}{\partial t}+\vv \cdot \nabla\vv\right)+\gamma\rho\vv=-\frac{1}{m} \rho \nabla \frac{\delta F[\rho]}{\delta \rho}.
\label{eq:xx_1}
\end{equation}
Dividing through by the fluid density we obtain:
\begin{equation}
\frac{D\vv}{Dt}+\gamma\vv=-\frac{1}{m} \nabla \frac{\delta F[\rho]}{\delta \rho} .
\label{eq:xx_3}
\end{equation}
where
\begin{equation}
\frac{D}{Dt} \equiv \frac{\partial }{\partial t}+\vv \cdot \nabla
\end{equation}
is the substantive derivative. Taken together, Eqs.\ \eqref{eq:xx_2} and \eqref{eq:xx_1} are one of the main results of this section of the paper. Given a suitable approximation for the free energy functional $F_{ex}[\rho]$ in \eqref{eq:F}, Eqs.\ \eqref{eq:xx_2} and \eqref{eq:xx_1} may then be simultaneously solved (with suitable boundary conditions) for the dynamics of the system.

It is worth noting at this point that when $\gamma$ is large, the first term on the left hand side of Eq.\ \eqref{eq:xx_3} is negligible with respect to the second term on the left hand side, giving
\begin{equation}
\gamma\vv \simeq -\frac{1}{m} \nabla \frac{\delta F[\rho]}{\delta \rho} .
\label{eq:xx_4}
\end{equation}
On substituting this expression into Eq.\ \eqref{eq:xx_2}, we obtain Eq.\ \eqref{eq:MT_DDFT_a}, the original DDFT of Marconi and Tarazona.\cite{MT,MT2}

In the opposite limit, when $\gamma \to 0$, Eq.\ \eqref{eq:xx_1} reduces to a generalised Euler equation:
\begin{equation}
m \rho\left(\frac{\partial \vv}{\partial t}+\vv \cdot \nabla\vv\right)=-\rho \nabla \frac{\delta F[\rho]}{\delta \rho}
\label{eq:euler}
\end{equation}
(recall that here we use the notation that $\rho(\rr,t)$ is the colloidal fluid number density). If we make a local density approximation (LDA) for the Helmholtz free energy functional \eqref{eq:F}:
\begin{eqnarray}
F[\rho]=\int \dr f(\rho(\rr,t))+\int \dr V^{ext}(\rr)\rho(\rr,t),
\label{eq:F_LDA}
\end{eqnarray}
where $f(\rho)=k_BT\rho (\ln \Lambda^3\rho-1)+f_{ex}(\rho)$ is the intrinsic Helmholtz free energy density, then from Eq.\ \eqref{eq:euler} we obtain
\begin{eqnarray}
m \rho\left(\frac{\partial \vv}{\partial t}+\vv \cdot \nabla\vv\right)&=&-\rho \nabla \frac{\partial f}{\partial \rho}-\rho \nabla V^{ext}
\end{eqnarray}
which may be rewritten in the following way:
\begin{eqnarray}
m \rho\left(\frac{\partial \vv}{\partial t}+\vv \cdot \nabla\vv\right)&=&-\nabla p-\rho \nabla V^{ext},
\label{eq:euler_2}
\end{eqnarray}
where $p=-f+\mu \rho$ is the local pressure and $\mu\equiv\partial f/\partial \rho$. At equilibrium, $\mu$ is the chemical potential. This equation is the Euler equation from Fluid Dynamics.\cite{Acheson,Batchelor}

We are now in a position to see the consequence of making the local equilibrium approximation \eqref{eq:MB_LE} for the one particle distribution function $f^{(1)}(\rr,\pp,t)$ in Eq.\ \eqref{eq:mainres}. This approximation leads to effectively setting the shear viscosity $\eta=0$ (recall that one obtains the Euler equation by setting $\eta=0$ in the Navier-Stokes equation). To go beyond this approximation, one may follow Kreuzer\cite{Kreuzer} and assume that the distribution function $f^{(1)}(\rr,\pp,t)$ can be expanded as a Taylor series, as follows:
\begin{eqnarray}
f^{(1)}(\rr,\pp,t)= f^{(1)}_{l.e.}(\rr,\pp,t)+a_1(|\pp-m\vv|)\left[\frac{(\pp-m \vv)\cdot \LL\cdot (\pp-m \vv)}{(\pp-m \vv)^2}-\frac{1}{3} \nabla \cdot \vv \right]+\cdots
\label{eq:MB_LE_extended}
\end{eqnarray}
where the function $a_1$ may also be a function of position and $\LL$ is the symmetric rate of strain tensor, whose components are
\begin{eqnarray}
\Lambda_{\alpha \beta}=\frac{1}{2}\left( \frac{\partial v_\alpha}{\partial r_\beta}+\frac{\partial v_\beta}{\partial r_\alpha}\right).
\end{eqnarray}
With this approximation in Eq.\ \eqref{eq:mainres} and further assuming that the one-body density of the colloids is a constant (i.e.\  that it is an incompressible fluid), we obtain an additional term in Eq.\eqref{eq:xx_1} that is $\simeq \frac{1}{m} \eta^{(K)} \nabla^2 \vv$, where $\eta^{(K)}$ is the kinetic energy contribution to $\eta$, the coefficient of shear viscosity, and is formally given by an integral over the distribution function $a_1(\pp)$.\cite{Kreuzer} Similar considerations at the two body level in the interaction terms in Eq.\ \eqref{eq:DDFT_1} give an addition contribution $\simeq \frac{1}{m} \eta^{(V)} \nabla^2 \vv$.\cite{Kreuzer} The resulting dynamical equation
\begin{equation}
m \rho\left(\frac{\partial \vv}{\partial t}+\vv \cdot \nabla\vv\right)+\gamma m\rho \vv=-\rho \nabla \frac{\delta F[\rho]}{\delta \rho} +\eta\nabla^2\vv,
\label{eq:NS}
\end{equation}
where $\eta=\eta^{(K)}+\eta^{(V)}$, is a generalisation of the Navier-Stokes equation for an incompressible fluid. For colloidal fluids, we believe that taking Eq.\ \eqref{eq:NS}  together with the continuity equation \eqref{eq:xx_2} may provide a basis for reliably describing the fluid dynamics, even when the density is not a constant (i.e.\ when the fluid is compressible). It should be noted, however, that although one may obtain formal expressions for the quantities $\eta^{(K)}$ and $\eta^{(V)}$ (see Ref.\ \onlinecite{Kreuzer} for further details), in practice it may be necessary to obtain this quantity by other means.

Finally in this section, we remind the reader that at equilibrium, where $\vv=0$ and $\partial \rho /\partial t=0$ all of the dynamical equations derived above yield the following expression:
\begin{equation}
\nabla \frac{\delta F[\rho]}{\delta \rho}=0,
\end{equation}
which may be integrated to give
\begin{equation}
\frac{\delta F[\rho]}{\delta \rho}=\mu,
\end{equation}
where $\mu$ is the chemical potential. This equation for the one-body density profile $\rho(\rr,t)$ is exact.\cite{HM, Evans92} Thus, given a reliable approximation for $F[\rho]$, all of the theories presented above exhibit the key feature that at equilibrium they yield the correct fluid one-body density profile.

\section{Dynamics of the one-body density and connections to MCT}
\label{sec:MCT}

One of the most appealing features of the original DDFT of Marconi and Tarazona,\cite{MT} Eq.\ \eqref{eq:MT_DDFT_a}, is that it gives a description of the fluid dynamics solely in terms of the one body density $\rho$, as opposed to the description given in the previous section involving both the density $\rho$ and the fluid velocity $\vv$. We now show how to obtain an approximate theory that just involves the fluid density $\rho$. We take the divergence of Eq.~(\ref{eq:DDFT_1}) and use Eq.\ (\ref{eq:continuity}) to eliminate the terms involving $\partial \jj/\partial t$ and $\jj$ to obtain:
\begin{eqnarray}
\frac{\partial^2 \rho(\rr_1,t)}{\partial t^2}
+\gamma\frac{\partial \rho(\rr_1,t)}{\partial t}-\nabla\cdot \AAA(\rr,t)=
\frac{k_BT}{m}\nabla^2\rho(\rr,t)
+ \frac{1}{m}\nabla \cdot \left[\rho(\rr,t)
\nabla V^{ext}(\rr,t)\right] \nonumber \\
+\frac{1}{m}\nabla \cdot \int \dr' 
\rho^{(2)}(\rr,\rr',t)\nabla v_2(\rr-\rr') \nonumber \\
+\frac{1}{m}\nabla \cdot \int \dr' \int \dr''
\rho^{(3)}(\rr_1,\rr_2,\rr,_3,t)\nabla v_3(\rr,\rr',\rr'') +\cdots.
\label{eq:abc}
\end{eqnarray}
So far, no approximations have been made and, taken together, Eqs.~(\ref{eq:abc}) and (\ref{eq:B}) are exact. If we make the approximation used earlier, where we assume that we may use Eq.\ \eqref{eq:grad_c1_many} as an approximation for the terms involving the non-equilibrium fluid pair, triplet and higher correlation functions, then we obtain:
\begin{equation}
\frac{\partial^2 \rho}{\partial t^2}
+\gamma \frac{\partial \rho}{\partial t}-\nabla \cdot \AAA(\rr,t)=
\frac{1}{m}\nabla \cdot \left[ \rho(\rr,t) \nabla \frac{\delta
F[\rho(\rr,t)]}{\delta \rho(\rr,t)} \right].
\label{eq:mainres_sec3_a}
\end{equation}
Since $\AAA(\rr,t) \sim \jj \otimes \jj$, when $\bar{\pp}/\sqrt{2mk_BT}$ is small, $\AAA(\rr,t)$ will also be small and we expect that in such circumstances this term may be either neglected or its influence incorporated into a renormalised $\gamma^*$ -- i.e.\ $\gamma$ is replaced by the renormalised $\gamma^*=\nu$, where $\nu=k_BT/mD$ and $D$ is the self diffusion coefficient.\cite{Archer17} Neglecting the term involving $\AAA(\rr,t)$ in Eq.\ \eqref{eq:mainres_sec3_a} gives
\begin{equation}
\frac{\partial^2 \rho}{\partial t^2}
+\gamma \frac{\partial \rho}{\partial t}=
\frac{1}{m}\nabla \cdot \left[ \rho(\rr,t) \nabla \frac{\delta
F[\rho(\rr,t)]}{\delta \rho(\rr,t)} \right].
\label{eq:mainres_sec3}
\end{equation}
The appealing feature of this equation is that it gives a theory for the fluid dynamics solely in terms of the fluid density $\rho$.

One circumstance where the term involving $\AAA(\rr,t)$ may be neglected is when the fluid density is high and the system is not too far from equilibrium (i.e.\ when the average local velocity $\vv$ is small). This situation was explored in Ref.\ \onlinecite{Archer17}, for the case when $\gamma=0$, where it was shown that one may derive the MCT (a theory for a density fluctuation correlation function), starting from Eq.\ \eqref{eq:mainres_sec3}. The argument presented in Ref.\ \onlinecite{Archer17} is entirely applicable to the present case. We do not repeat the full argument here, merely reminding the reader of the salient points. One key issue is that we now interpret the one body density $\rho$ as a coarse grained density field. For example, we may follow Refs.\ \onlinecite{Archer8} and \onlinecite{Archer17} and define the temporally coarse grained density $\rho(\rr,t)=\int_{-\infty}^{\infty}\dd t K(t-t') \hat{\rho}(\rr,t')$, where $\hat{\rho}(\rr,t)=\sum_{i=1}^N\delta(\rr-\rr_i(t))$ is the density operator (recall that $\rr_i(t)$ is the location of the $i$th particle at time $t$), and $K(t)$ is a normalised function of finite support. The precise shape of $K(t)$ defines the degree of coarse graining. The coarse grained density $\rho(\rr,t)$ exhibits thermal fluctuations and the amplitude of these depend on the extent of the coarse graining.\cite{Archer8} The dynamical equations for the coarse grained density depend on a coarse grained two-body density distribution function $\rho^{(2)}(\rr,\rr',t)=\int_{-\infty}^{\infty}\dd t K(t-t') \hat{\rho}(\rr,t')\hat{\rho}(\rr',t')$ (and in the case when $v_3\neq0$ it also depends on a similarly defined three-body distribution function). If we then assume that we have coarse grained sufficiently that Eq.\ \eqref{eq:grad_c1_many} still holds true for this coarse grained $\rho^{(2)}(\rr,\rr',t)$, then following the argument presented above, we obtain Eq.\ \eqref{eq:mainres_sec3} as the equation governing the time evolution of the coarse grained one-body density profile.

The exact excess Helmholtz free energy functional \eqref{eq:F}, that is required as input to Eq.\ \eqref{eq:mainres_sec3}, is unknown. However, we may approximate this quantity by Taylor expanding in powers of $\delta \rho(\rr,t) =\rho(\rr,t)-\rho_b$, where $\rho_b$ is the bulk density of the uniform fluid. Truncating the expansion beyond terms of ${\cal O}(\delta \rho^2)$ we obtain:\cite{Evans92,Archer17}
\begin{eqnarray}
F_{ex}[\rho(\rr,t)] = F_{ex}[\rho_b] -c^{(1)}(\infty)
\int \dr \delta\rho(\rr,t) 
- \frac{k_BT}{2} \int \dr \int \dr'
\delta{\rho}(\rr,t)\delta{\rho}(\rr',t) c^{(2)}(\rr-\rr'),
\label{eq:F_ex}
\end{eqnarray}
where
\begin{equation}
c^{(2)}(\rr,\rr')=\beta \frac{\delta^2 F_{ex}[\rho]}{\delta \rho(\rr) \delta \rho(\rr')} 
\end{equation}
is the pair direct correlation function.\cite{Evans92,HM} On substituting Eq.~(\ref{eq:F_ex}) into Eq.~(\ref{eq:mainres_sec3}) and then Fourier transforming, we obtain:\cite{Archer17}
\begin{eqnarray}
\ddot{\rho}_{\kk}(t)+\gamma \dot{\rho}_{\kk}(t)
=-\frac{k^2}{\beta m} \rho_{\kk}(t)+\frac{\rho_b k^2}{\beta m}
c_{\kk} \rho_{\kk}(t)
+\frac{1}{\beta m}\frac{1}{(2 \pi)^3} \int \dk' \kk\cdot \kk'
\rho_{\kk'}(t)c_{\kk'}\rho_{\kk-\kk'}(t),
\label{eq:Spin_dec_eq_2_next}
\end{eqnarray}
where $c_{\kk}$ is the Fourier transform of $c^{(2)}(r)$ and $\rho_{\kk}(t)$ is the Fourier transform of
$\delta \rho(\rr,t)$. We multiply through in Eq.~(\ref{eq:Spin_dec_eq_2_next}) by $\rho_{-\kk}(0)$, and then average over the ensemble of initial configurations of the density field. Making the approximation proposed in Ref.\ \onlinecite{Archer17} (see also Ref.\ \onlinecite{Kawasaki95}) for the quantity $\langle \hat{R}_{\kk}(t)\hat{R}_{-\kk}(0)\rangle$, where
\begin{eqnarray}
\hat{R}_{\kk}(t)=\frac{1}{\beta m (2 \pi)^3} \int \dk' \kk \cdot \kk'
\rho_{\kk'}(t)c_{\kk'}\rho_{\kk-\kk'}(t),
\label{eq:R_hat}
\end{eqnarray}
we obtain the MCT equation
\begin{eqnarray}
\ddot{\phi}_{\kk}(t)+\gamma \dot{\phi}_{\kk}(t)
+\Omega_{\kk}^2 \phi_{\kk}(t)
=-\int_0^t \dd t' m_{\kk}(t')\dot{\phi}_{\kk}(t-t'),
\label{eq:MCT_eq}
\end{eqnarray}
for the normalised density fluctuation correlation function
\begin{eqnarray}
\phi_{\kk}(t)=\frac{\langle \rho_{\kk}(t)\rho_{-\kk}(0)\rangle}
{\langle \rho_{\kk}(0)\rho_{-\kk}(0)\rangle},
\label{eq:correlator}
\end{eqnarray}
where $\Omega_{\kk}^2=k^2/\beta m S_{\kk}$ and $S_{\kk}=[1-\rho_b c_{\kk}]^{-1}$ is the static structure factor. Making the usual MCT approximation of factorising four--point correlation functions into products of two--point correlation functions one obtains the standard MCT
expression for the memory function:\cite{HM,Archer17,Gotze}
\begin{eqnarray}
m_{\kk}(t)=\frac{k_BT \rho_b}{2 (2 \pi)^3k^2m}
\int \dk'\left[\kk\cdot\kk'c_{\kk'}+\kk \cdot(\kk-\kk')c_{\kk-\kk'}\right]^2
S_{\kk'}S_{\kk-\kk'}\phi_{\kk'}(t)\phi_{\kk-\kk'}(t).
\label{eq:m}
\end{eqnarray}
The MCT obtained above for under-damped colloidal particles formally has the same structure as the standard MCT that is used to elucidate the properties of the glass transition in atomic/molecular fluids\cite{HM,Gotze} -- i.e.\ systems with deterministic Newtonian dynamics -- see Eq.\ \eqref{eq:eqs_of_motion_not_damped}. What the above analysis shows is that results obtained for atomic and molecular fluids\cite{HM,Gotze} are also relevant to under-damped colloidal suspension. Since the dynamics in dense glassy systems is strongly influenced by how {\em collective} density fluctuations decay, it is perhaps not surprising that the decay of $\phi_\kk(t)$ should be governed by equations having a structure which do not strongly depend on the underlying equations of motion for the {\em individual} particles. Note also that for overdamped colloids (large $\gamma$), we may neglect the term $\partial^2 \rho/\partial t^2$ from Eq.\ \eqref{eq:mainres_sec3}, to obtain Eq.\ \eqref{eq:MT_DDFT_a}. On following the argument presented above, starting from Eq.\ \eqref{eq:MT_DDFT_a}, we obtain a MCT equation the same as Eq.\ \eqref{eq:MCT_eq}, but with the term $\ddot{\phi}_{\kk}(t)$ omitted from the left hand side. This MCT for overdamped colloids was originally obtained by Szamel and L\"owen.\cite{SzamelLowenPRA1991}

\section{Summary and concluding remarks}
\label{sec:conc}

In this paper we have derived a number of approximate DDFTs, Eqs.\ \eqref {eq:xx_1}, \eqref{eq:NS} and \eqref{eq:mainres_sec3}, for systems of Brownian particles where inertial effects are relevant. These theories are also relevant to atomic and molecular fluids, which correspond to the limit $\gamma \to 0$ in these equations. To derive these DDFTs we made two approximations. The approximation that pertains to all three cases is to assume that one may use Eq.~(\ref{eq:grad_c1_many}), which strictly only applies at equilibrium. In using this, we effectively assume that the two-body spatial correlations in the non-equilibrium fluid are equal to those in an equilibrium fluid with the same one body density profile. As discussed in Refs.\ \onlinecite{MT3} and \onlinecite{MarconiMelchionna}, this is essentially a mean field approximation that is reliable only in the high friction limit or for systems interacting via pair potentials that are continuous and differentiable. In Refs.\ \onlinecite{MT3} and \onlinecite{MarconiMelchionna} the authors consider the particular case of a system of hard spheres. The collision dynamics of such a system is somewhat different to that in systems of particles interacting via potentials that vary continuously. They show that to describe the hard-sphere collision effects, one must approximate the term on the right hand side of Eq.\ \eqref{eq:mom_bal_1} involving the two-body distribution function by a term that contains a binary collision operator\cite{MarconiMelchionna} (the three-body potential is zero). Such an approach may also prove to be useful for extending the theory presented here.

The second approximation that was made here was to assume a particular form for the phase space probability distribution function $f^{(1)}$ -- see Eqs.\ \eqref{eq:MB_LE} and \eqref{eq:MB_LE_extended}. Making approximations at this level are justified for systems that are not too far from equilibrium. However, for very strongly driven systems, one must expect to have to go beyond such approximations to obtain a reliable description of the fluid dynamics.

The approach presented here provides a fully microscopic basis for the equations used to consider systems with diffuse interfaces in fluid mechanics. Either one-component (gas-liquid) systems may be considered,\cite{Andersonetal} or the present results may easily be extended to consider a two component fluid, in order to study the dynamics of a system exhibiting fluid-fluid phase separation.\cite{Andersonetal, JasnowVinals, Uwe} These approaches may be obtained by assuming a gradient expansion of the Helmholtz free energy\cite{Evans92,HM} in the present theory. As mentioned in the introduction, and we emphasise again here: the DDFTs presented here are microscopic generalisations of the Euler and Navier-Stokes equations. Much is known in the literature about the solutions of these equations.\cite{Acheson, Batchelor} One such method worth mentioning, due to having some underlying connections to the present approach, is the Lattice-Boltzmann method.\cite{SwiftetalPRL1995, Yeomans2006} The present work shows how to build upon this knowledge and to incorporate into these theories information about the microscopic structure and correlations in the fluid via the Helmholtz free energy functional.

A further extension of the theory presented in this paper is to use the DDFT to obtain a MCT theory that is applicable for studying the glass transition in colloidal suspensions in which inertial effects in the particle dynamics are important. To derive the MCT, we assume that we may neglect the term $\nabla \cdot \AAA(\rr,t)$ in Eq.~(\ref{eq:mainres_sec3_a}). This term is significant when currents are large (i.e.\ when the condition $\vv \ll \sqrt{k_BT/m}$ no longer holds) and so makes the theory unreliable in such cases. However, for dense fluids that are near to equilibrium or with a low average velocity $\vv$, one should find that the DDFT in Eq.\ \eqref{eq:mainres_sec3} is relevant. We should also mention that stochastic dynamical equations, that are of a similar structure to the dynamical equations that we have obtained here (Eqs.\ \eqref {eq:xx_1}, \eqref{eq:NS}), have been developed to study the dynamics in glassy systems.\cite{DasRMP2004, DasetalPRL1985, DasMazenkoPRA1986, MiyazakietalPRE2004} Within the present DDFT framework, these stochastic analogues of Eqs.\ \eqref {eq:xx_1} and \eqref{eq:NS} may be viewed as having been obtained by considering the dynamics of a coarse grained density field, along the lines presented in Ref.\ \onlinecite{Archer8}.

We should also make a few further comments concerning Eq.\ \eqref{eq:mainres_sec3}. Equations of this general form have appeared before in the literature in a number of different contexts. When the free energy functional $F[\rho]$ in Eq.\ \eqref{eq:mainres_sec3} is set to be simply that of an ideal gas (i.e.\ when $F_{ex}[\rho]=0$) and we set the external potential $V^{ext}(\rr)=0$ in Eq.\ \eqref{eq:F}, then we obtain
\begin{equation}
\frac{\partial^2 \rho}{\partial t^2}
+\gamma \frac{\partial \rho}{\partial t}=
\frac{k_BT}{m}\nabla^2 \rho.
\label{eq:telegrapher}
\end{equation}
This equation is telegrapher's equation and has numerous applications. Equations of this general form have been used for example in the description of heat waves,\cite{JosephPreziosiRMP1989} electric currents \cite{WeissPhysicaA2002} and nuclear collision dynamics.\cite{AzizGavinPRC2004} When $F[\rho]$ is replaced by the Ginzburg-Landau free energy functional, a generalised Cahn-Hilliard equation is obtained. Such an equation was proposed by Galenko and co-workers as a phase-field model for binary alloys \cite{GalenkoPLA2001, GalenkoJouPRE2005, GalenkoLebedevPML2007, GalenkoLebedevPLA2008} and also by Koide {\it et.~al.} \cite{KoideetalPLB2006, KoideetalBJP2007} as a means of incorporating memory: if one assumes overdamped particle dynamics (Eq.\ \eqref{eq:eqs_of_motion_damped}) and a noise field ${\bf G}_i(t)$ with memory -- i.e.\  where $\left< \xi_i^{\alpha}(t)\xi_i^{\nu}(t')\right> \propto \delta_{ij} \delta^{\alpha \nu} \exp(-|t-t'|/\tau)$, then following the argument of Koide {\it et.~al.}, one obtains an equation of the same form as Eq.\ \eqref{eq:mainres_sec3}.

To conclude, we recall that the present theory is for systems with underlying equations of motion given by Eq.~(\ref{eq:eqs_of_motion}). Therefore, the DDFT will not be reliable for systems where the equations of motion are not well-modelled by Eq.~(\ref{eq:eqs_of_motion}), such as in colloidal fluids where hydrodynamic interactions between the colloids are significant. One may include the hydrodynamic interactions by extending the original Marconi-Tarazona DDFT \eqref{eq:MT_DDFT_a} to include additional terms that describe the hydrodynamic interactions at the Rotne-Prager level.\cite{RexLowen} Alternatively, one may perhaps be able to use the present DDFTs to incorporate hydrodynamic effects by treating such systems as two component mixtures. The hydrodynamic interactions between the colloids would enter the treatment via the density field of the second (solvent) species.

I gratefully acknowledge helpful discussions with Lubor Frastia, Markus Rauscher and Uwe Thiele and thank RCUK for financial support.



\begin{thebibliography}{99}

\bibitem{Kreuzer}
H.J. Kreuzer, {\it Nonequilibrium Thermodynamics and its Statistical
Foundations}, New York: Oxford University Press, (1981).

\bibitem{Acheson}
D.J. Acheson, {\it Elementary Fluid Dynamics}, Oxford University Press (1990).

\bibitem{Batchelor}
G.K. Batchelor, {\it An Introduction to Fluid Dynamics}, Cambridge University Press (1967).

\bibitem{microfluidics}
See for example T.M. Squires and S.R. Quake, Rev. Mod. Phys. {\bf 77}, 977 (2005) and references therein.

\bibitem{tweezers}
See for example J.E. Molloy and M.J. Padgett, Contemporary Phys. {\bf 43}, 241 (2002) and references therein.

\bibitem{MT}
U. Marini Bettolo Marconi and P. Tarazona, J. Chem. Phys. {\bf 110}, 8032 (1999).

\bibitem{MT2}
U. Marini Bettolo Marconi and P. Tarazona, J. Phys.: Condens. Matter {\bf 12}, A413 (2000).

\bibitem{Dhont_book}
J.K.G. Dhont,  {\it An Introduction to Dynamics of Colloids}, Elsevier, Amsterdam, (1996).

\bibitem{Archer7}
A.J. Archer and R. Evans, J. Chem. Phys. {\bf 121}, 4246 (2004).

\bibitem{Archer11}
A.J. Archer, J. Phys.: Condens. Matter {\bf 17}, 1405 (2005).

\bibitem{footnote1}
Note that $\rho(\rr,t)$ is the ensemble average density, i.e. the average over all realisations of the stochastic noise. For more details on this point see Refs.\ \onlinecite{MT}, \onlinecite{MT2} and \onlinecite{Archer8}.

\bibitem{Archer8}
A.J. Archer and M. Rauscher, J. Phys. A: Math. Gen. {\bf 37}, 9325 (2004).

\bibitem{Evans92}
R. Evans, in {\it Fundamentals of Inhomogeneous Fluids}, ed. D. Henderson, Dekker, New York, (1992), ch. 3.

\bibitem{HM}
J.-P. Hansen and I.R. McDonald, {\it Theory of Simple Liquids}, Academic, London, (2006), 3rd ed.

\bibitem{joe:christos}
J. Dzubiella and C.N. Likos, J. Phys.: Condens. Matter {\bf 15}, L147 (2003).

\bibitem{flor2}
F. Penna and P. Tarazona, J. Chem. Phys. {\bf 68}, 1766 (2003). 

\bibitem{flor1}
F. Penna, J. Dzubiella and P. Tarazona, Phys. Rev. E {\bf 68}, 061407 (2003).

\bibitem{Archer13}
A.J. Archer, J. Phys.: Condens. Matter {\bf 17}, S3253 (2005).

\bibitem{RexetalPRE2005}
M. Rex, H. L\"owen and C.N. Likos, Phys. Rev. E {\bf 72}, 021404 (2005).

\bibitem{RexetalMolP2006}
M. Rex, C.N. Likos, H. L\"owen and J. Dzubiella, Mol. Phys. {\bf 104}, 527 (2006).

\bibitem{Matthiasetal}
C.P. Royall, J. Dzubiella, M. Schmidt, A. Van Blaaderen, Phys. Rev. Lett. {\bf 98}, 188304 (2007).

\bibitem{MT3}
U. Marini Bettolo Marconi and P. Tarazona, J. Chem. Phys. {\bf 124}, 164901
(2006).

\bibitem{MarconiMelchionna}
U.M.B. Marconi and S. Melchionna, J. Chem. Phys. {\bf 126} 184109 (2007).

\bibitem{Archer17}
A.J. Archer, J. Phys.: Condens. Matter {\bf 18}, 5617 (2006).

\bibitem{Kawasaki95}
K. Kawasaki, Transp. Theory Stat. Phys. {\bf 24}, 755 (1995).

\bibitem{Gotze}
W. G\"otze, {\it Liquids, Freezing and Glass Transition} ed. J.-P. Hansen, D. Levesque and J. Zinn-Justin (Amsterdam: North Holland) (1991).

\bibitem{vanKampen}
N.G. van Kampen, {\it Stochastic Processes in Physics and Chemisty}, North Holland, Amsterdam, (1990), 6th ed.

\bibitem{HM86}
J.-P. Hansen and I.R. McDonald, {\it Theory of Simple Liquids}, Academic, London, (1986), 2nd ed.

\bibitem{SzamelLowenPRA1991}
G. Szamel and H. L\"owen, Phys. Rev. A {\bf 44}, 8215 (1991).

\bibitem{Andersonetal}
D.M. Anderson and G.B. McFadden and A.A. Wheeler, Annu. Rev. Fluid Mech. {\bf 30}, 139 (1998).

\bibitem{JasnowVinals}
D. Jasnow and J. Vi\~nals, Phys. Fluids, {\bf 8}, 660 (1996).

\bibitem{Uwe}
U. Thiele, S. Madruga and L. Frastia, Phys. Fluids {\bf 19}, 122106 (2007).

\bibitem{DasRMP2004}
S.P. Das, Rev. Mod. Phys. 76, {\bf 785} (2004).

\bibitem{DasetalPRL1985}
S.P. Das, G.F. Mazenko, S. Ramaswamy, and J. Toner, Phys. Rev. Lett. {\bf 54}, 118 (1985).

\bibitem{DasMazenkoPRA1986}
S.P. Das and G.F. Mazenko, Phys. Rev. A {\bf 34}, 2265 (1986).

\bibitem{MiyazakietalPRE2004}
K. Miyazaki, D.R. Reichman, R. Yamamoto, Phys. Rev. E, {\bf  70} 011501 (2004).

\bibitem{SwiftetalPRL1995}
M.R. Swift, W.R. Osborn and J.M. Yeomans, Phys. Rev. Lett. {\bf 75}, 830 (1995).

\bibitem{Yeomans2006}
J.M. Yeomans, Physica A {\bf 369}, 159 (2006).

\bibitem{JosephPreziosiRMP1989}
D.D. Joseph and L. Preziosi, Rev. Mod. Phys. {\bf 61}, 41 (1989); D.D. Joseph and L. Preziosi, Rev. Mod. Phys. {\bf 62}, 375 (1990).

\bibitem{WeissPhysicaA2002}
G.H. Weiss, Physica A, {\bf 311} 381 (2002).

\bibitem{AzizGavinPRC2004}
M.A. Aziz and S. Gavin, Phys. Rev. C {\bf 70}, 034905 (2004).

\bibitem{GalenkoPLA2001}
P. Galenko, Phys. Lett. A {\bf 287}, 190 (2001).

\bibitem{GalenkoJouPRE2005}
P. Galenko and D. Jou, Phys. Rev. E {\bf 71}, 046125 (2005).

\bibitem{GalenkoLebedevPML2007}
P. Galenko and V. Lebedev, Philosophical Magazine Letters {\bf 87}, 821 (2007).

\bibitem{GalenkoLebedevPLA2008}
P. Galenko and V. Lebedev, Phys. Lett. A {\bf 372}, 985 (2008).

\bibitem{KoideetalPLB2006}
T. Koide, G. Krein and R.O. Ramos, Phys. Lett. B {\bf 636}, 96 (2006).

\bibitem{KoideetalBJP2007}
T. Koide, G. Krein and R.O. Ramos, Braz. J. Phys. {\bf 37}, 601 (2007).

\bibitem{RexLowen}
M. Rex and H. L\"owen, Phys. Rev. Lett. {\bf 101}, 148302 (2008).

\end{thebibliography}
\end{document}